\documentclass{iaus}
\usepackage{graphicx}

\title[Symp. 265~~Chemo-Dynamical History of the Milky Way] 
{The Chemo-Dynamical History of the Milky Way as Revealed by SDSS/SEGUE}

\author[T.C. Beers]   
{Timothy C. Beers$^{1,2}$}

\affiliation{$^1$Department of Physics \& Astronomy, Michigan State
University, \break  email: beers@pa.msu.edu \\[\affilskip]
$^2$Joint Institute for Nuclear Astrophysics}

\pubyear{2009}
\volume{265}  
\jname{Chemical Abundances in the Universe: Connecting First Stars to Planets}
\editors{K. Cunha, M. Spite \& B. Barbuy, eds.}
\begin{document}

\maketitle

\begin{abstract}

Although originally conceived as primarily an extragalactic survey,
the Sloan Digital Sky Survey (SDSS-I), and its extensions SDSS-II and
SDSS-III, continue to have a major impact on our understanding of the
formation and evolution of our host galaxy, the Milky Way. The
sub-survey SEGUE: Sloan Extension for Galactic Exploration and
Understanding, executed as part of SDSS-II, obtained some 3500 square
degrees of additional $ugriz$ imaging, mostly at lower Galactic
latitudes, in order to better sample the disk systems of the Galaxy.
Most importantly, it obtained over 240,000 medium-resolution spectra
for stars selected to sample Galactocentric distances from 0.5 to 100
kpc. In combination with stellar targets from SDSS-I, and the recently
completed SEGUE-2 program, executed as part of SDSS-III, the total
sample of SDSS spectroscopy for Galactic stars comprises some 500,000
objects.

The development of the SEGUE Stellar Parameter Pipeline has enabled
the determination of accurate atmospheric parameter estimates for a
large fraction of these stars. Many of the stars in this data set
within 5 kpc of the Sun have sufficiently well-measured proper motions
to determine their full space motions, permitting examination of the
nature of much more distant populations represented by members that
are presently passing through the solar neighborhood. Ongoing analyses
of these data are being used to draw a much clearer picture of the
nature of our galaxy, and to supply targets for detailed
high-resolution spectroscopic follow-up with the world's largest
telescopes. Here we discuss a few highlights of recently completed and
ongoing investigations with these data.  


\end{abstract}

\firstsection 
\section{Introduction}

The era of the massive surveys of Galactic stars is now very much
underway. Our understanding of the history of the formation and
evolution of the stellar populations in the Galaxy is presently being
revolutionized, based on results from two primary surveys, the Sloan
Digital Sky Survey (SDSS; \cite[York et al. 2000]{York00}), in
particular its dedicated sub-surveys Sloan Extension for Galactic
Exploration and Understanding (SEGUE-1 and SEGUE-2; \cite[Yanny et al.
2009]{Yan09}) and the RAdial Velocity Experiment (RAVE; \cite[Zwitter
et al. 2008]{Zwi08}; Wyse, this volume).
Collectively, these two surveys are in the process of providing
detailed spectroscopic information for a million or more stars,
sampling all of the known stellar populations in the Galaxy. Although
refinements in the procedures for extracting estimates of the stellar
atmospheric parameters (T$_{\rm eff}$, log g, [Fe/H]) and individual element
abundances (in the case of SDSS/SEGUE, [$\alpha$/Fe] and [C/Fe] 
ratios; for RAVE, many more) are still underway, the wealth of
information already available provides the basis for numerous
investigations and follow-up observations.\footnote{The most recent
public release for SDSS/SEGUE is DR-7, which includes SEGUE-1; see
\cite[Abazajian et al. (2009)]{Aba09}. The next public release for 
SDSS/SEGUE is DR-8, scheduled for December 2010, which will include results
from SEGUE-2.}  Here we summarize a few of the ongoing projects that are being enabled 
by the SDSS/SEGUE database.

\begin{figure}[t]
\begin{center}
\includegraphics[width=4.0in]{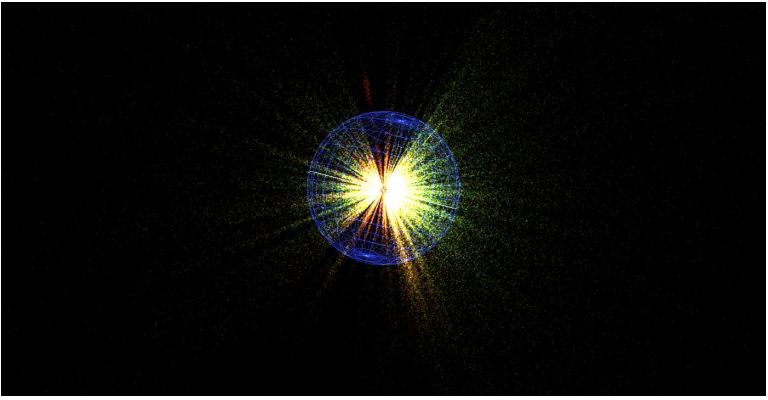}
\caption{Simple visualization of the SDSS/SEGUE stellar database, for
roughly 400,000 stars with available atmospheric parameter estimates
from the SSPP.  The sphere has a radius of 10 kpc, centered on the
Sun.  The colors scale with metallicity, from green (metal-poor stars) to 
red (metal-rich stars).}
\label{fig1}
\end{center}
\end{figure}

\section{The SEGUE Stellar Parameter Pipeline and a Visualization of
the SDSS/SEGUE Stellar Database}

The SEGUE Stellar Parameter Pipeline (SSPP) processes the wavelength-
and flux-calibrated spectra generated by the standard SDSS
spectroscopic reduction pipeline, obtains equivalent widths and/or
line indices for about 80 atomic or molecular absorption lines, and
estimates the effective temperature, T$_{\rm eff}$, surface gravity,
log g, and metallicity, [Fe/H], for a stellar spectrum through the
application of a number of approaches. A given method is usually
optimal over specific ranges of color and signal-to-noise ($S/N$)
ratio. The SSPP employs 8 primary methods for the estimation of
T$_{\rm eff}$, 10 for the estimation of log g, and 12 for the
estimation of [Fe/H]. The final estimates of the atmospheric
parameters are obtained by robust averages of the methods that are
expected to perform well for the color and $S/N$ of the spectrum
obtained for each star. The use of multiple methods allows for
empirical determinations of the internal errors for each parameter,
based on the range of reported values from each method -- typical
internal errors for stars in the temperature range that applies to the
calibration stars are $\sigma_{\rm Teff} \sim$ 100 K to $\sim$ 125 K,
$\sigma_{\rm log g} \sim $ 0.25 dex, and $\sigma_{\rm [Fe/H]}
\sim$ 0.20 dex. The external errors in these determinations are of similar size. See
\cite[Lee et al. (2008a)]{Lee08a}, \cite[Lee et al. (2008b)]{Lee08b},
and \cite[Allende Prieto et al. (2008)]{All08} for more details.

Over the past several years, large-aperture telescopes have been used
to obtain high-resolution spectroscopy for over 300 of the brighter
(14.0 $< g_0 <$ 17.0) SDSS stars (see \cite[Allende Prieto et al.
2008]{All08}; Aoki et al., in preparation; Lai et al., in
preparation). The observations reported by Aoki et al. and Lai et al.
suggest that the current SSPP is actually somewhat conservative in the
assignment of metallicities for stars of the lowest [Fe/H], in the
sense that high-resolution estimates of [Fe/H] are on the order of 0.3
dex lower than those reported by the SSPP.

Based on the derived atmospheric parameters from the SSPP, and
distances estimated from photometric parallaxes, Figure 1 shows a
simple visualization of the SDSS/SEGUE database. The colors are coded
to represent metallicity, and clearly indicate the presence of stars
from the disk populations close to the Galactic plane, and an extended
population of stars comprising representatives of the inner- and
outer-halo populations. More sophisticated visualizations, encoding
other available observables such as full space motions, are presently
being developed.

\section{The Impact of SDSS/SEGUE on Searches for Metal-Poor Stars}

The great majority of metal-poor stars in the Galaxy identified to
date have come from two primary sources, the HK survey of Beers and
colleagues (\cite[Beers et al. 1985]{Bee85}; \cite[Beers et al.
1992]{Bee92}) and the Hamburg/ESO Survey of Christlieb and
colleages (\cite[Christlieb et al. 2008]{Chr08}). SDSS/SEGUE
specifically targeted large numbers of low-metallicity candidates
based on combinations of SDSS filters that approximately isolate them
from the vastly more common higher metallicity stars of the disk
populations. This approach, although not as selective as prism-survey
techniques, has greatly enlarged the numbers of known metal-poor
stars, as summarized in Table 1. It should be kept in mind that, due
to the ongoing calibration of the SSPP at lower metallicities, the
true numbers of extremely metal-poor stars identified by SDSS/SEGUE
with [Fe/H] $< -3.0$ is certain to increase in the near future. The
apparent lack of newly discovered ultra ([Fe/H] $< -4.0$) and hyper
([Fe/H] $< -5.0$) metal-poor stars by SDSS/SEGUE could also be, at
least in part, due to calibration issues. In order to be
certain, all stars with metallicities suggested by the SSPP to be
below [Fe/H] $= -3.0$ need to be observed at higher spectral
resolution, as the presence of interstellar Ca~II K (the only metallic
feature detectible at such low metallicities from medium-resolution
spectra) can confound the derived abundance as well.

\begin{table}[h]
  \begin{center}
  \caption{Numbers of known metal-poor stars pre- and post-SDSS/SEGUE}
  \label{tab1}
 {\scriptsize
  \begin{tabular}{|l|c|r|r|}\hline 
{\bf Metallicity Class} & {\bf $<$ [Fe/H]} & {\bf N (Pre-SDSS)} & {\bf N (Post-SDSS)} \\ \hline
Metal-Poor              &  $< -1.0$        &    15,000          &  150,000+ \\
Very Metal-Poor         &  $< -2.0$        &     3,000          &   30,000+ \\
Extremely Metal-Poor    &  $< -3.0$        &       400          &    1,000+ \\
Ultra Metal-Poor        &  $< -4.0$        &         5          &        5  \\
Hyper Metal-Poor        &  $< -5.0$        &         2          &        2  \\
Mega Metal-Poor         &  $< -6.0$        &         0          &        0  \\ \hline
\end{tabular}
}
\end{center}
\end{table}

\begin{figure}[th]
\begin{center}
\includegraphics[width=3.0in]{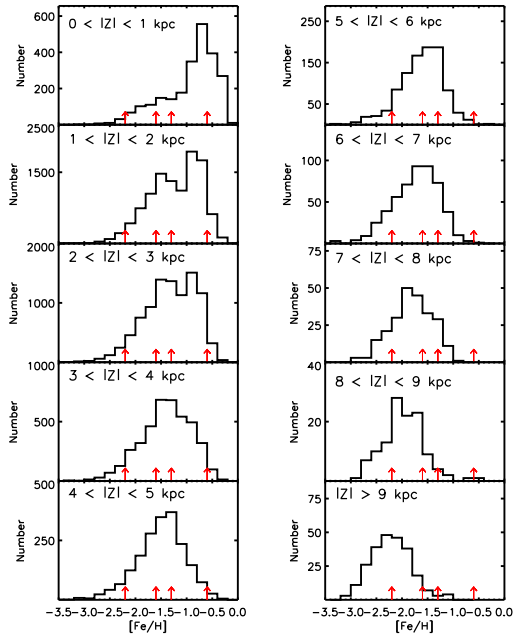}
\vspace{-25pt}
\caption{Observed metallicity distribution functions (MDFs) for the full sample
of SDSS/SEGUE DR-7 calibration stars as a function of vertical distance from the
Galactic plane. The black histograms represent the MDFs obtained at different
cuts of $|$Z$|$, while the red arrows denote the locations of the metallicity
peaks of the MDF for the thick disk ($-0.6$), the MWTD ($\sim -1.3$), the inner
halo ($-1.6$), and the outer halo ($-2.2$), respectively.}
\label{fig2}
\end{center}
\end{figure}

\section{An Update on the Inner/Outer-Halo Dichotomy}

Based on an analysis of a large sample of calibration stars
from SDSS DR-5 (\cite [Adelman-McCarthy et al.
2007]{Ade07}), \cite[Carollo et al. (2007)]{Car07} argued for the
existence of {\it at least} a two-component halo. In their view, the
Galactic halo comprises two broadly overlapping structural components,
an inner and an outer halo. These components exhibit different spatial
density profiles, stellar orbits, and stellar metallicities. It was
found that the inner-halo component dominates the population of halo
stars found at distances up to 10-15 kpc from the Galactic center,
while the outer-halo component dominates in the region beyond 15-20
kpc. The inner halo was shown to comprise a population of stars
exhibiting a flattened spatial density distribution, with an inferred
axial ratio on the order of 0.6. According to these authors,
inner-halo stars possess generally high orbital eccentricities, and
exhibit a small (or zero) net prograde rotation around the center of
the Galaxy. The metallicity distribution function (MDF) of the inner
halo peaks at [Fe/H] = $-$1.6, with tails exending to higher and lower
metallicities. By comparison, the outer halo comprises stars that
exhibit a more spherical spatial density distribution with an axial
ratio $\sim$ 0.9. Outer-halo stars possess a wide range of
eccentricities, exhibit a clear retrograde net rotation, and are drawn
from an MDF that peaks at [Fe/H] = $-$2.2, a factor of four lower than
that of the inner-halo population.

This work has now been extended by \cite[Carollo et al. (2009)]{Car09} (see
also Carollo's article, this volume), making use
of the calibration star sample through DR-7.  The full sample of
32360 stars represents an increase of 60\% relative to the numbers
used in the previous analysis.  The larger sample, which includes
many more stars from the disk populations due to SEGUE observations,
has been used to derive, among many other results, estimates of the
velocity ellipsoids for the inner- and outer-halo populations, as well
as for the metal-weak thick disk (MWTD) and canonical thick-disk populations.

One can obtain an impression of the contribution of the various
populations to the observed MDF from inspection of Figure 2, which
shows its variation with height above the Galactic plane. Examination
of the left-hand column of panels in this figure shows how the MDF
changes from the upper panel, in which there are obvious contributions
from the thick-disk, MWTD, and inner-halo components in the cuts close
to plane, to the lower panel, with an MDF dominated by inner-halo
stars. In the right-hand column of panels, with distances from the
plane greater than 5 kpc, the transition from inner-halo dominance to
a much greater contribution from outer-halo stars is obvious. This
demonstration is, by design, independent of any errors that might
arise from derivation of the kinematic parameters, and provides
confirmation of the difference in the chemical properties of the
inner- and outer-halo populations originally suggested by
\cite[Carollo et al. (2007)]{Car07}.

Although it might appear possible to attempt mixture-model analysis to obtain
MDFs for each of the individual components, we recall that biases in the
selection of the stars in the calibration-star sample would confound such an
attempt. Other samples of SDSS stars are being studied for this
purpose, and will be reported on in due course.

\section{ECHOS in the Inner Halo}

\cite[Schlaufman et al. (2009)]{Sch09} have used the SEGUE 
spectroscopic database of metal-poor main-sequence turnoff (MPMSTO)
stars to search for significant radial velocity enhancements along 137
SEGUE-1 lines of sight. They identify ten (seven for the first time)
Elements of Cold Halo Substructure (ECHOS) in the volume within 17.5
kpc of the Sun, in the inner halo of the Milky Way. These ECHOS
represent the observable stellar debris of ancient merger events in
the stellar accretion history of the Milky Way. 

These authors use their detections and completeness estimates to infer
a formal upper limit of $0.34 \pm 0.2$ on the fraction of the MPMSTO
population in the inner halo that belong to ECHOS. They also suggest
that there exists a significant population of low fractional
overdensity ECHOS in the inner halo, and predict that one-third of the 
inner halo (by volume) harbors ECHOS with MPMSTO number densities
$n \approx 15$ kpc$^{-3}$. In addition, they estimate that there are
$\approx$ 10$^3$ ECHOS in the entire inner halo. Since ECHOS are likely
older than the known substructures identified by surface-brightness
contrast methods, these detections provide a direct measure of the
accretion history of the Milky Way in a region and time interval that
has yet to be fully explored. The authors argue, on the basis of this
information, that the level of merger activity has been roughly
constant over the past few Gyrs, and that there has been no accretion
of stellar systems more massive than a few percent of the Milky
Way mass in that interval.

\section{[$\alpha$/Fe] Ratios for SDSS/SEGUE Stars}

Lee et al. (in preparation) demonstrate, by comparison with measured
abundances of [$\alpha$/Fe] from high-resolution spectra in the ELODIE
spectral library and recently obtained $R = 15000$ spectra from
Hobby-Eberly Telescope observations of SDSS/SEGUE stars, that it is
possible to determine [$\alpha$/Fe] ratios from SDSS spectra (with
$S/N > 20$) to an accuracy of better than 0.1 dex, for stars with
atmospheric parameters in the range T$_{\rm eff}$ = [4500,7500] K, log
g = [1.5,5.0], and [Fe/H] = [$-$2.5,$+$0.2]. This capability opens up
the opportunity to make use of this ratio to investigate
predictions from simulation studies of the formation and evolution of 
the disk systems of the Milky Way (e.g., \cite[Abadi et al.
2003]{Aba03}; \cite[Brook et al. 2007]{Bro07}; \cite[Kazantzidis et
al. 2008]{Kaz08};  \cite[Ro\v skar et al. 2008]{Rov08}; 
\cite[Sch\"onrich \& Binney 2009a]{Sch09a}; \cite[Sch\"onrich \&
Binney 2009b]{Sch09b}), as well as of the halos of the
Galaxy (e.g., \cite[Johnston et al. 2008]{Joh08}).  

\begin{figure}[th]
\begin{center}
\includegraphics[width = 3.0in]{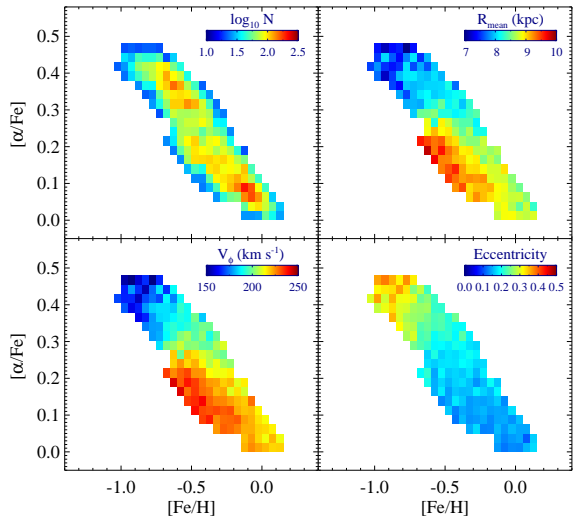}
\caption{Distribution of number densities, mean orbital radii, 
rotation velocities, and orbital eccentricities 
for F- and G-type dwarfs within 2 kpc of the Galactic plane, located
between $8 < \rm R < 9$ kpc, in bin sizes of 0.05 dex in [Fe/H] by
0.025 dex in [$\alpha$/Fe].  $R_{\rm mean}$ is an average of $R_{\rm max}$ and
$R_{\rm min}$, the maximum and minimum projected distance on the plane
reached by a star over the course of its orbit. Each pixel has more
than 20 stars. }
\label{fig3}
\end{center}
\end{figure}

For example, the lower right panel of Figure 3 shows, for a sample of
some 10000 SDSS/SEGUE F- and G-dwarfs in the solar neighboorhood, that
stars with higher [$\alpha$/Fe] ratios (associated with the thick
disk) have orbital eccentricities that peak around 0.2, with
relatively few such stars having higher eccentricities (most of these
are found at lower [Fe/H], indicating possible membership in the
MWTD population). The association of the low-metallicity,
high-[$\alpha$/Fe] stars with the MWTD population is
also indicated by the lower left panel of Figure 3, where it is clear
that they exhibit a substantially lower mean rotational velocity. Much
remains to be explored with data such as these.

\section{[C/Fe] Ratios for SDSS/SEGUE Stars}

Stars with observed carbon enhancements are likely to play a key role 
in our understanding of the formation and evolution of the various
stellar populations in the Galaxy, as the production mechanisms of
carbon depend sensitively on the initial mass function and star
formation environments of their birth.  Although carbon-enhanced stars
were not specifically selected for in SDSS/SEGUE, they are found in great
numbers over a variety of the categories that were targeted, such as metal-poor
stars, F-turnoff stars, and late-type giants. 

\begin{figure}[h]
\begin{center}
\includegraphics[width=2.52in, height=2.5in]{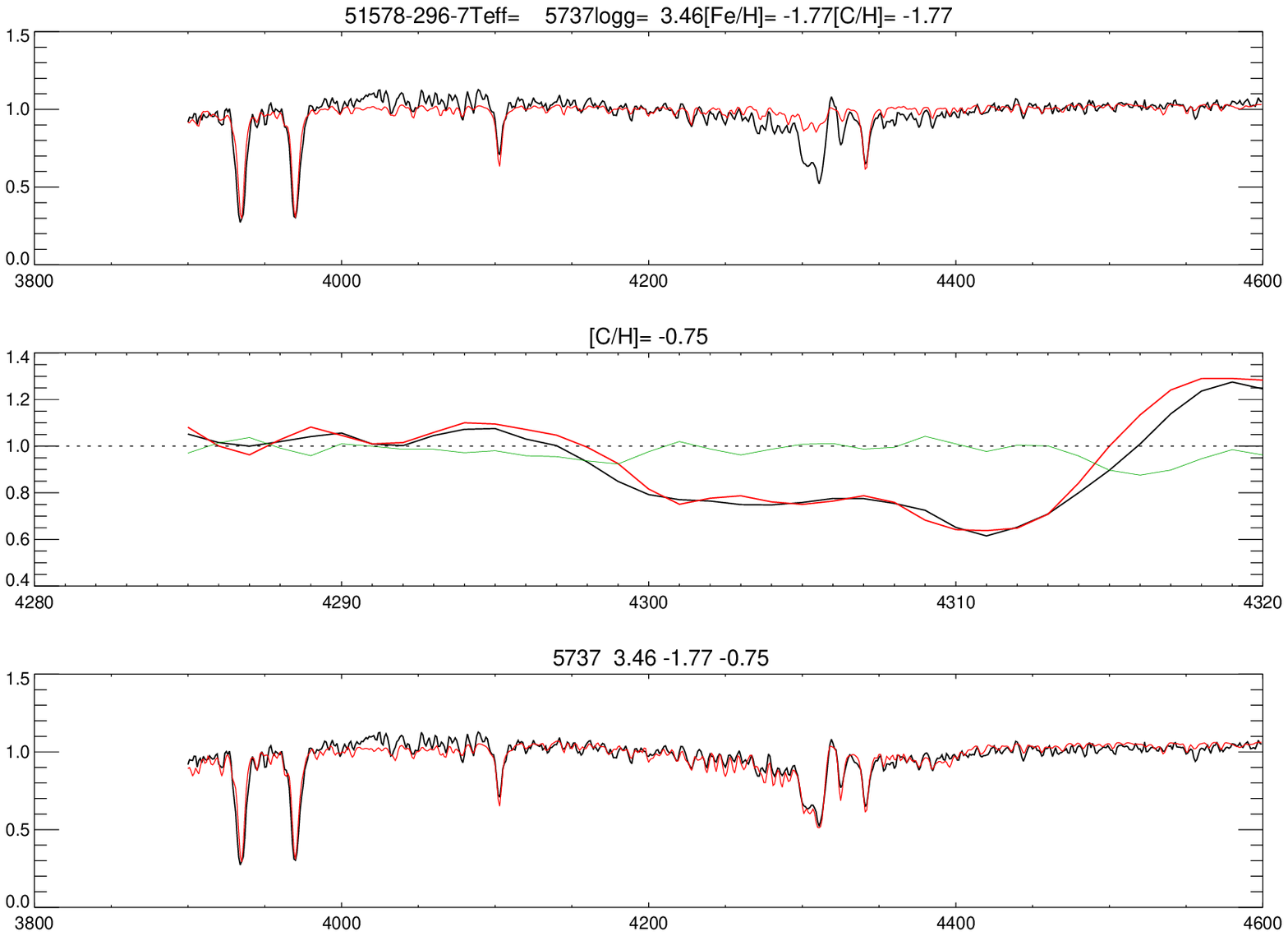}
\includegraphics[width=2.52in, height=2.5in]{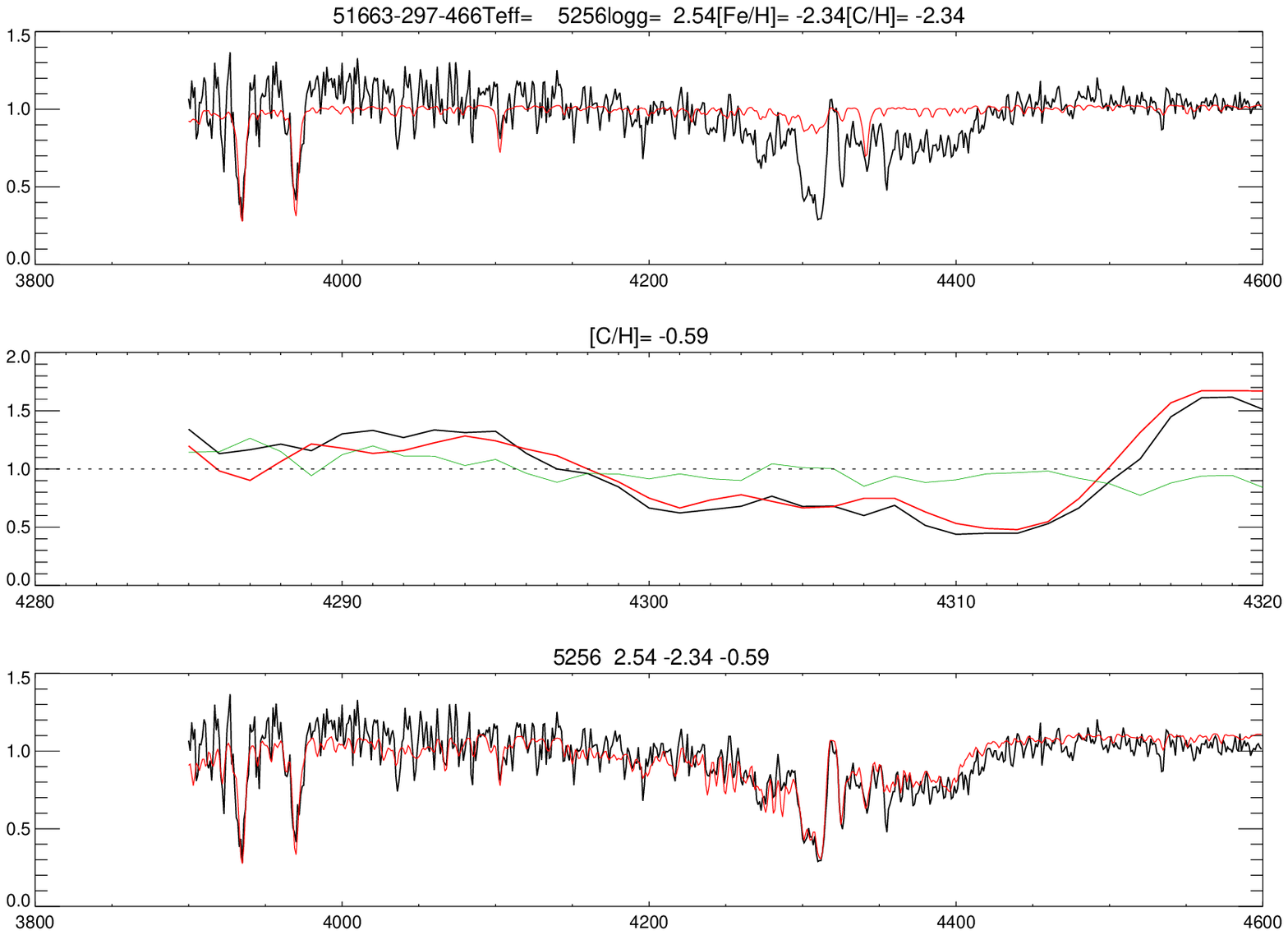} 
\caption{Two examples of the fitting procedure for determination of
[C/Fe] in metal-poor stars.  The upper panels show a portion of
the original spectrum (black) overlayed with a synthetic spectrum
(red) with the listed parameters and [C/Fe] = 0. The middle panels
show the region around the CH G-band, with a red line showing the best
fit. The green line is a division of the original spectrum by the fit
spectrum, which should be close to 1.0 for a successful fit. The lower
panels show the result of the best fit with the listed [C/H] level{\it
Left}: A moderately metal-poor star. {\it Right}: A very metal-poor
star.}
\label{fig4}
\end{center}
\end{figure}

Over the course of the past few years, members of the SEGUE team have
been working to develop and refine methods for estimation of [C/Fe]
from the SDSS spectra, with the goal of providing this information for
as many stars as possible in the upcoming DR-8 public release. For
spectra with sufficient S/N (roughly $> 15/1$) these methods are now
able to estimate [C/Fe], or upper limits on [C/Fe], accurate to roughly
0.1 dex. See Figure 4 for some examples of fits to the CH G-band
region, which is the feature used to estimate [C/Fe].  
There are several hundred thousand stars in SDSS/SEGUE for
which such estimates now exist.  Based on these data, Sivarani et al.
(in preparation) is obtaining a definitive estimate of the
much-debated frequency of Carbon-Enhanced Metal-Poor (CEMP) stars, and
in particular, the dependence of this fraction on [Fe/H] between 
solar and [Fe/H]$ = -3.0$.  Carollo et al. (in preparation) is
examining whether or not the fraction of CEMP stars differs among
the various disk and halo stellar populations identified in the
SDSS/SEGUE calibration stars, which has been argued by \cite[Tumlinson
(2007)]{Tum07} to probe the influence of the CMB on the IMF at early times. 

\section{Subdwarf M stars in SDSS/SEGUE}

\cite[Lepine (2009)]{Lep09} and collaborators have been actively pursuing techniques
for the identification and analysis of subdwarf M stars based on SDSS
colors, and have demonstrated that they can be usefully
separated into metallicity sub-classes, which they refer to as
``subdwarfs" (sdM), ``extreme subdwarfs" (esdM), and ``ultra subdwarfs"
(usdM), in order of decreasing metal content. Such stars were
specifically targeted in SEGUE-2. As a result of this follow-up, there
are now more than 6000 such stars known, an order of magnitude
increase in the numbers that were known just a few years ago. 

\begin{figure}[t]
\begin{center}
\includegraphics[width = 2.52in]{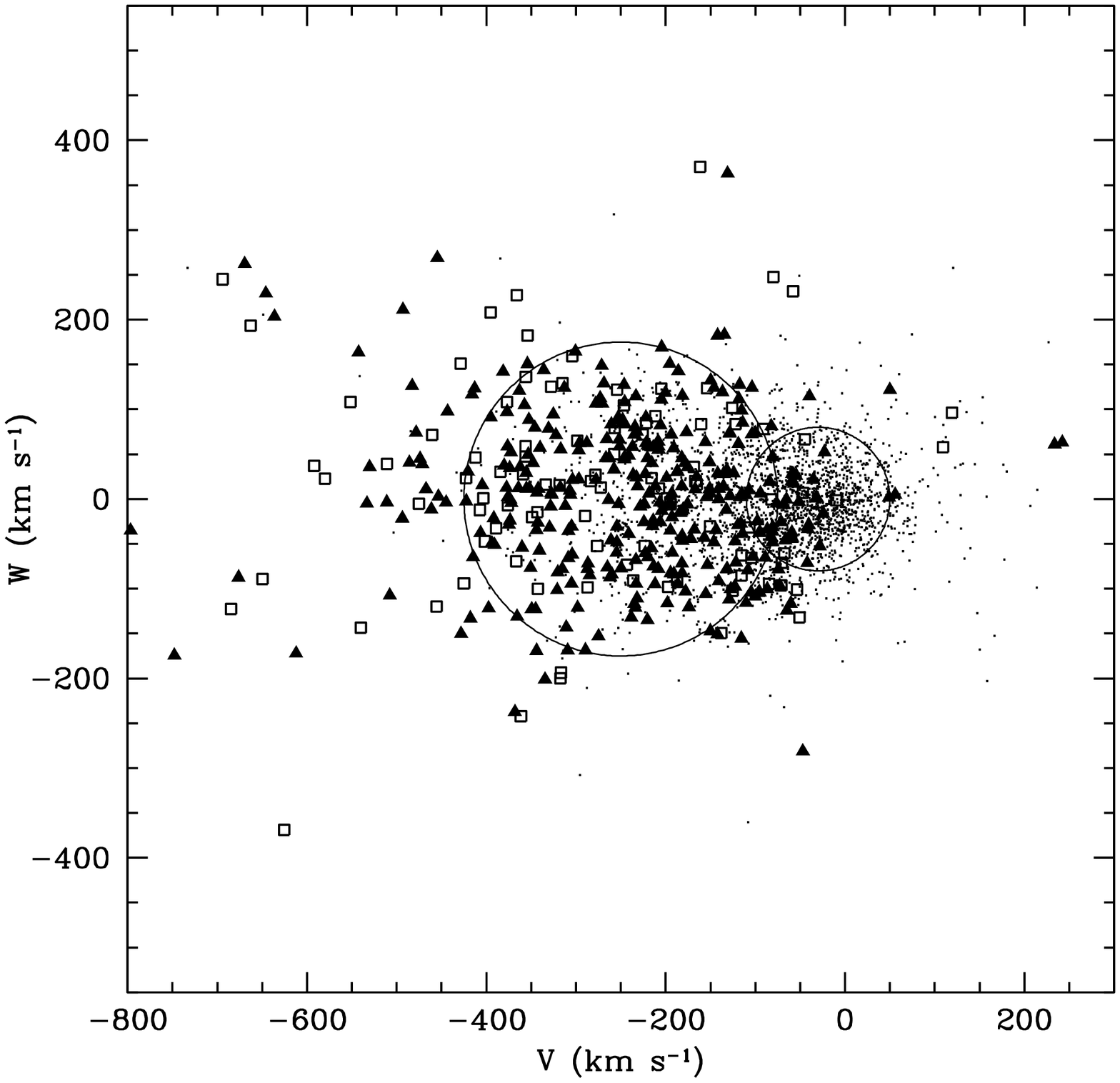}
\includegraphics[width = 2.52in]{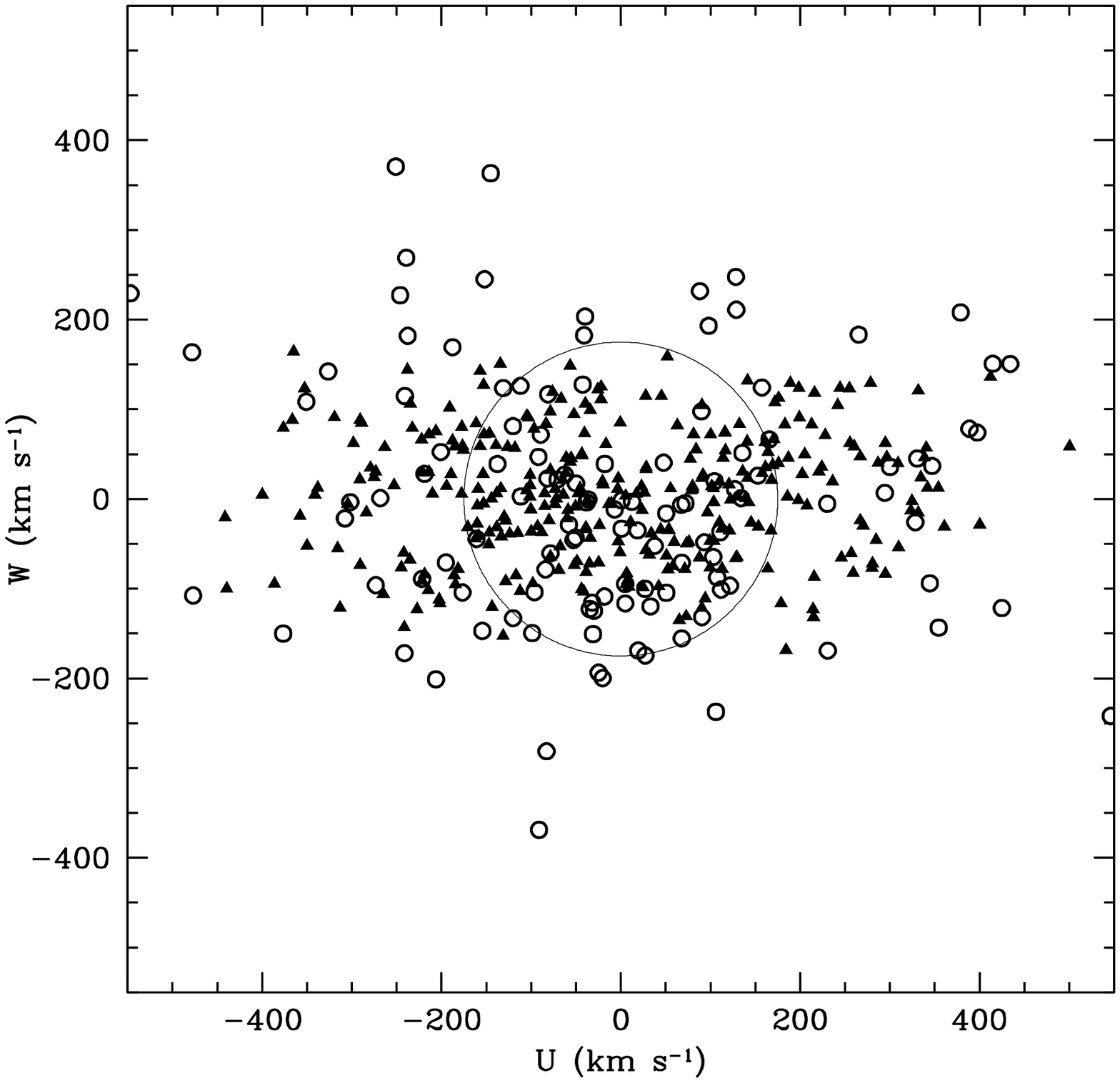}
\caption{{\it Left:} Distribution of $W$ vs. $V$ velocities for the sample of sdM
stars (small dots), esdM stars (filled triangles), and usdM stars
(open squares) from SDSS/SEGUE.  The smaller circle indicates the
centroid for the sdM stars, while the larger circle is that for the
esdM stars. Stars with a velocity component $V > -$220 km~s$^{-1}$ are
on prograde orbits, while those with $V > -$220 km~s$^{-1}$ are on
retrograde orbits. The tendency for the usdM stars to occupy orbits
with large retrograde rotations is clear. {\it Right}: Distribution of
$W$ vs. $U$ velocities for esdM stars (filled triangles) and usdM
stars (open circles) stars. Note the higher energy orbits associated with the usdM
stars, suggesting membership in the outer-halo population. }
\label{fig5}
\end{center}
\end{figure}

Because sdM, esdM, and usdM stars are selected in a completely
different way than other probes in SDSS/SEGUE, they provide a useful
comparison to kinematic analyses of, e.g., the calibration stars
studied by \cite[Carollo et al. (2007)]{Car07} and \cite[Carollo et al.
(2009)]{Car09}. It is thus of great interest to note, as shown in
Figure 5, that the kinematics of this very local sample reflects the 
dichotomy of the halo that has been previously claimed. As can be
appreciated from inspection of this figure, the sdM stars (the
centroid indicated by the smaller circle) have local kinematics that
can be associated with the thick disk, the esdM stars (the centroid
indicated by the large circle) with the inner-halo population, and of
greatest interest, the tendency of the usdM stars to exhibit a net
retrograde rotation, which can be associated with the outer-halo
population.  

   
Funding for SDSS-III has been provided by the Alfred P. Sloan
Foundation, the Participating Institutions, the National Science
Foundation, and the U.S. Department of Energy. The SDSS-III web site
is http://www.sdss3.org/.

SDSS-III is managed by the Astrophysical Research Consortium for the
Participating Institutions of the SDSS-III Collaboration including the
University of Arizona, the Brazilian Participation Group, University
of Cambridge, University of Florida, the French Participation Group,
the German Participation Group, the Michigan State/Notre Dame/JINA
Participation Group, Johns Hopkins University, Lawrence Berkeley
National Laboratory, Max Planck Institute for Astrophysics, New Mexico
State University, New York University, the Ohio State University,
University of Portsmouth, Princeton University, University of Tokyo,
the University of Utah, Vanderbilt University, University of Virginia,
University of Washington and Yale University. 
    
{}

\end{document}